# A Review of Research on Civic Technology: Definitions, Theories, History and Insights

Weiyu Zhang, Department of Communications and New Media, National University of Singapore
Gionnieve Lim, Singapore University of Technology and Design
Simon Perrault, Singapore University of Technology and Design
Chuyao Wang, Hong Kong University of Science and Technology

There have been initiatives that take advantage of information and communication technologies to serve civic purposes, referred to as civic technologies (Civic Tech). In this paper, we present a review of 224 papers from the ACM Digital Library focusing on Computer Supported Cooperative Work and Human-Computer Interaction, the key fields supporting the building of Civic Tech. Through this review, we discuss the concepts, theories and history of civic tech research and provide insights on the technological tools, social processes and participation mechanisms involved. Our work seeks to direct future civic tech efforts to the phase of "by the citizens".

CCS Concepts: • **Human-centered computing** • **Human-centered computing ~ Collaborative and social computing ~ Collaborative and social computing theory, concepts and paradigms ~ Computer supported cooperative work**

**KEYWORDS:** civic technology; civic engagement; citizen participation; democratic participation

**ACM Reference format:**

First author, Second author, Third author. 2021. A Review of Research on Civic Technology: Definitions, Theories, History and Insights. *ACM Comput. Surv.*

## 1  INTRODUCTION

Starting from the late 20th century, Information and Communication Technologies (ICTs) became a driving force that revolutionized matters regarding the civics, including both governments and civil societies. Governments were the first to develop and apply ICTs in their daily operations, often characterized as computerization and digitization in the 1980-1990s [1]. The 21st century witnessed the emergence of society-led initiatives to take advantage of ICTs to serve civic purposes. As ICTs developed from infrastructure to applications, the space in which citizens and their collectivities are able to work became larger. Technologists founded Hacker and Maker communities to socialize; out of these communities, civically-driven technologists organized projects that aim for the public good (e.g., Code for America was founded in 2009 [37]). Since a Knight Foundation report pronounced the term "civic technology" in 2013 [25], ICTs and the civics are becoming mutually dependent. We cannot define the civic without referring to the technology that supports/disrupts it; neither can we afford imagining technology without taking the civic into account.

   Both public services offered by the government and ground-up initiatives by the citizens now have to function through the technology. Government portals and mobile applications replaced offices and plain mails to serve the citizens. Social media were used by citizens to express opinions, organize actions, and form collectivities. Although the technological development has been changing the world rapidly, recent years witnesses that our imagination of technology may have neglected the civic perspective. From social media platforms selling private data to artificial intelligence modelled after human biases, highlighting the civic considerations in technological development becomes immediately urgent.

This understanding of civic tech in practice has been broad [76] and include both government-centric and citizen-centric approaches. The former approach focuses on enabling governments to provide service and engage citizens in their policymaking. The latter approach emphasizes the empowerment of citizens, who not only interact with governments but also connect and collaborate with each other. Moreover, the latter approach often centers on "digital initiatives by civil society, private organizations, and individual citizens" [79]. A large number of civic tech initiatives exists. Skarzauskiene and Maciuliene [79] found 614 such platforms and Saldivar and colleagues' [73] keyword search returned with 1,246 counts. The Civic Tech Field Guide[1] records more than 4,000 tech for good projects.

The field of Civic Tech is international, cross-sectoral, and interdisciplinary. From its inception, Civic Tech has been internationalized – the UK government was the first to establish its Digital Service, a dedicated unit to develop and implement citizen-faced technologies now found in many other governments [1]; the US was the birthplace for the "code for all" groups, now spreading all over the world including the African continent [37,47,75]. Asia has not only joined the global trend but also played defining roles in shaping this international movement. Taiwan's g0v initiative[2], for instance, represents a unique model of integrating grassroots and government-led efforts in developing civic tech. Civic Tech is not just an academic field and it captures the imagination of practitioners from both the public and private sectors. One of the earliest mentions of Civic Tech comes from a Knight Foundation report [25]. Big tech companies (e.g., IBM's Call for Code Global Challenge), social enterprises (e.g., Brussel-based CitizenLab.co), governments (e.g., Singapore's GovTech), and organizations (e.g., UK's non-profit organization mysociety.org) have all actively promoted Civic Tech. As an academic field, the interdisciplinary nature of Civic Tech research is found in the overlapping interests in developing technologies for civic purposes among scholars from communication, computer science, information science, political science, and public administration. Although not all scholars build computer technologies, they do contribute to understanding the mechanisms behind how the technologies work, as well as attempting to provide purely social, and more often, hybrid (social plus technological) designs to address civic challenges.

Acknowledging that Civic Tech is a topic that captures a wide range of interests, this paper focuses on one particular group of interests: academic works mainly in the interdisciplinary fields of Computer Supported Cooperative Work (CSCW) and Human-Computer Interaction (HCI) documented in the Association of Computing Machinery (ACM) Digital Library. The reason behind this is that these two communities are the key academic force that supports the building of civic tech. Through this comprehensive review of civic tech in the two closely related fields (CSCW and HCI), we aim to provide a piece of work to describe, analyze, and provide constructive suggestions for Civic Tech studies. While a pioneering review casted a wide web (1,246 in initial set) but caught a relatively small number of fishes (35 in final set) [73], our ACM-focused review tries to achieve a balance between scope and depth. Our contributions are as follows:

---

[1] https://civictech.guide/
[2] https://www.theguardian.com/world/2020/sep/27/taiwan-civic-hackers-polis-consensus-social-media-platform

- This paper clarifies basic concepts and theories regarding civic tech, so future works can be equipped with these accumulated knowledge;
- This paper provides a longitudinal description about the history of the civic tech research filed, pointing out both achievements and inadequacies;
- This paper evaluates the technological tools, the social processes, and the participation mechanisms used by prior studies, directing future civic tech works to the phase of "by the citizens".
- This paper provides an open-source dataset of 224 papers with our applied qualitative codes and the metadata, enabling future meta-analyses by the civic tech community.

## 1.1 Related Works

Reflecting on the field of civic tech requires an expansive understanding that goes beyond the ACM full papers. The history of public usage of information and communication technologies (ICTs) can be dated back to the 1980s when governments started to digitalize their operations [1]. Back then, the introduction of computers, especially the desktops, enabled a digital transformation inside governments. From records, documents, to communication, these digital tools help government employees to take advantage of the digital affordances such as storage space and longevity. When the diffusion of ICTs reached a large proportion of the society in early 2000s, e-government emerged as a new phase of digitalization that focused on interacting with citizens through ICTs [28]. International Federation for Information Processing (IFIP) launched its first eGOV conference in 2001. UNESCO offered a definition of e-governance in 2011 as "the public sector's use of ICTs with the aim of improving information and service delivery, encouraging citizen participation in the decision-making process and making government more accountable, transparent, and effective." This e-governance phase is accompanied by another important non-technological development, which is the dwindling public interest in political participation across the world and thus, a concerted cry for political innovations.

The wide adoption of ICTs, particularly the Internet, has led to vivid discussions on how democracies or citizen engagement can be used to implement the political innovations such as participatory democracy or deliberative democracy. Governments set up e-consultation websites or online feedback channels. Scholars debated on what roles citizens can play in policy-making processes, other than being consulted. The IFIP e-PART conference was launched in 2008, and a range of e- or digital democracy entities hold academic events (e.g., CeDEM). Among various political innovations, deliberation [22] or minipublic [44] or participatory budgeting [15] happened to be on the rise around the same period. Technologists quickly turned the deliberation model online, creating a sub-field called "online deliberation" [13,46,50,87].

The era of social media or what used to be called Web 2.0 allows ordinary users to create and share data at a fast speed. This technological shift changed both e-government and e-participation. Governance has to extend beyond government-owned online portals and merge into the social networks sustained through social media. Participation, on the other hand, seemed to become both easy when everyone can post their opinions on social media and difficult when users only hear from their echo chambers. It has become evident that the commercial nature of social media drove the platforms to profit from gathering users' private data and manipulating users' behaviors. These commercial platforms have also become barriers for other ICT tools, including both e-government and e-participation ones, to reach out to their users [27].

The civic tech scholarship this review paper tries to cover mostly comes from a period that either parallels or postdates the domination of social media since the 2010s. A field scan via interviews with technologists was published in 2012 [60]. The Knight Foundation report was released in 2013 [25]. Other efforts to review appeared in book chapters [32,76] and a non-ACM journal [73]. Our historical analysis later will show that the field of civic tech has matured to a stage when a sufficient number of studies emerged, and the body of scholarship shows a clear distinction from prior e-gov or e-part research. A high-level meta-review thus becomes necessary.

## 2 METHOD

### 2.1 Data Collection

To build the review corpus, the ACM Digital Library was used as it contains the most comprehensive collection of records covering the fields of computing and information technology with over 2 million full-text publications[3]. A search was conducted in the ACM Full-Text collection with the criteria that the article contained the word 'civic' in 'Title', 'Abstract' and 'Author Keyword' and was published prior to 26 July 2021, the time of the search. We looked specifically into 'Conferences' and 'Journals'. In total, the search returned 496 articles. We excluded articles that had 7 pages or less as these were usually not full papers to reach 239 articles. After reading all the abstracts, we removed 12 papers that had the keyword 'civic' but were not research regarding civic tech. We later removed 3 papers for varying reasons to reach the final corpus of 224 articles (Figure 1).

The keyword 'civic' in 'Author Keyword', 'Title' and 'Abstract' was used as it indicates articles in which the authors themselves distinctly identify 'civic' to be an important aspect of their article, enabling us to obtain highly relevant results as compared to casting a wide net for articles in which 'civic' was found in the full text. As mentioned above, e-government, e-participation and e-democracy studies have a long history, but they do not always embrace a civic perspective or emphasize benefiting citizens. We did not use the 'tech*' keyword for three reasons. Firstly, most authors did not consider 'tech*' a keyword even though, for example, their articles investigated the use of specific technologies. Secondly, 'tech*' is such a generic term that the produced search results were mostly irrelevant. And lastly, given that the source is the ACM Digital Library, we find that 'tech*' becomes a redundant keyword as most articles, even if in subtle ways, discuss technology.

---

[3] https://libraries.acm.org/digital-library

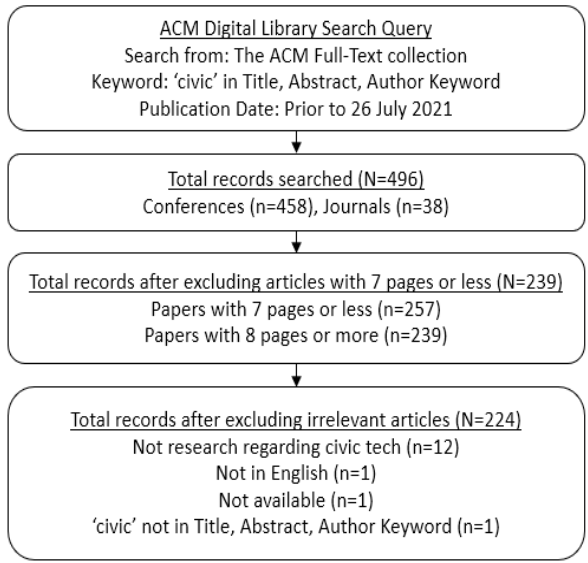

Figure 1. Corpus building process.

## 2.2 Content Analysis

The codebook was developed in three steps: firstly, the research team consulted prior review papers on civic tech [73,79] and related topics [59,65] in order to generate a set of initial codes; secondly, the research team ran a pilot test on the codebook by applying the codes on 50 full papers. The codebook was revised to reach consistency and increase readability. Finally, initial results were shared with experts in a CSCW 2020 workshop, Civic Technologies: Research, Practice and Open Challenges [3]. The codebook was further revised in light of the workshop participants' suggestions and concerns. The final codebook (see Appendix 1) was structured to generate data that can answer our queries about definitions of civic tech, theories used, publication features (e.g., venues, years, methods), and who used what tools through what process in what contexts.

The coding of the set of 224 articles took place from Aug 11, 2021, to Oct 21, 2021. Two graduate students, one taking a PhD degree in Computer Science and one taking a master's degree in Social Science, were trained to perform the coding work. The training process was completed on a random subset of 24 articles where intercoder reliabilities were calculated using ReCal[4], an online utility for computing intercoder reliability. Through a three-round training process, disagreements were discussed and reconciled, and the coding process was iteratively formalized. When the pairwise agreement using Krippendorff's alpha reached an acceptable level, the coders proceeded to each code half of the remaining articles. The IRR ranged from .61 to 1.0 and was on average .81 (SD=.12). The completed codes were then assessed to identify inconsistent results that were sent back for review, giving rise to the final set of codes (see Table 1).

---

[4] http://dfreelon.org/utils/recalfront/

Table 1. Code categories.

| Category | Codes | IRR | Selection |
|---|---|---|---|
| Definitions and theories | search the articles with "defin*" and "concept*" and "theor*" to record instances | NA | Multiple |
| Civic issues | Public service and governance; Urban planning; Environment; Education; Crime and safety; Accessibility and minority; News and journalism; Disasters; Neighborhood/local issues; Democracy and participation; N/A; Others | M=0.91 SD=0.00 | Single |
| Stakeholder (A) | Activists/advocates; Volunteers; Children/youths/adolescents; Disadvantaged groups; Administrators/government employees; Professional/occupational groups; Community members; Organisational members; General people; Students; N/A; Others | M=0.76 SD=0.31 | Multiple |
| Stakeholder (B) | Online; Offline; Both; N/A | M=0.91 SD=0.00 | Single |
| Methodology | Quantitative; Qualitative; Mixed method; N/A | M=1.00 SD=0.00 | Single |
| Data collection method | Interview; Survey; Observation; Data logging; Web scraping; Experiment; Focus group; Case study; Workshop; Review and reflection; N/A; Others | M=0.86 SD=0.14 | Multiple |
| Data source | Literature; Human; Physical device; Dataset; Purely virtual technology; N/A; Others | M=0.78 SD=0.37 | Multiple |
| Data analysis method | Thematic analysis; Content analysis; Modelling and simulation; Inferential statistics; Descriptive statistics; N/A; Others | M=0.74 SD=0.17 | Multiple |
| Study location (A) | Online; N/A | M=0.82 SD=0.00 | Single |
| Study location (B) | USA; America (exclude USA); UK; Europe (exclude UK); Asia; Africa; Australia and New Zealand; Global; N/A | M=1.00 SD=0.00 | Multiple |
| Type of civic tech (A) | Tailor-made tech; Existing platform; Hybrid; N/A | M=0.61 SD=0.00 | Single |
| Type of civic tech (B) | Social networking site or forum; Algorithm; Non-conventionally based; Mobile-based; Computer/web-based; Not specified; N/A; Others | M=0.61 SD=0.31 | Multiple |
| Civic tech owners | Academics; Companies; Governments; Communities; Organisations; General citizens; Not specified; N/A; Others | M=0.63 SD=0.33 | Multiple |
| Design method | Iterative design; User-centered design; Participatory design; Value sensitive design; Speculative Design; Co-design; N/A; Others | M=0.86 SD=0.33 | Multiple |
| Funder | School; Government; Foundation; Company; Offline community group; Online community group; Organisation; Network; N/A; Others | M=0.74 SD=0.41 | Multiple |
| Partner | School; Government; Foundation; Company; Offline community group; Online community group; Organisation; Network; N/A; Others | M=0.88 SD=0.32 | Multiple |

## 3. DEFINITIONS OF CIVIC TECH

### 3.1 What is Civic Tech?

Many definitions of civic tech exist: the Knight Foundation report [25] defined it as technology "promoting civic outcomes". As Schrock [76] observed, there has been a wide range of understanding about how technology can be used "for the public good" [84] or the "common good" [32]. The divergence of understanding hinges upon the definition of "civic" or what is considered "the good". We searched for explicit definitions or conceptualizations of civic tech in our corpus and found multiple instances. Figure 2 categorizes the found definitions of the civic into three layers. The understanding of technology is less controversial, although with a developing trend to extend our definition of technology to process and design.

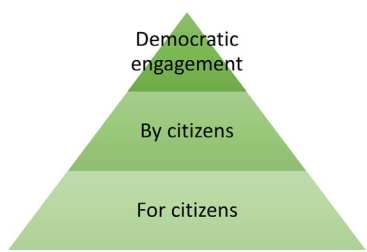

Figure 2. Definition of the notion "Civic".

### 3.2 Definitions of "Civic"

The divergent definitions of "civic" can be broken down into three layers: The most basic layer of the definition (as illustrated in Figure 2 as the bottom of the pyramid) emphasizes that the technology has to be for the benefits of citizens, regardless of who has a say in what is beneficial or good for citizens. The second layer of the definition (as illustrated in Figure 2 as the middle layer of the pyramid) adds more to benefiting citizens by including at least some engagement (in contrast to passive consumption) by citizens. While the first and second layers can be observed in all kinds of political systems, the top layer of the pyramid further refines the civic as bounded within democratic systems. We put this layer of definitions on the tip because democratic engagement by citizens is a narrower category than generic engagement by citizens.

*3.2.1 For Citizens.* The bottom layer in Figure 2 is present in all civic tech definitions but there are differences in what is considered "good" for citizens. The simplest version of "civic as good for citizens" is found in studies that address fundamental issues in citizens' everyday life such as water main breaks [49] and fire risks [89]. Solving these issues is self-evidently of benefit to citizens who reside in the affected areas, so a justification of why it is good for citizens is often absent in such studies. Moving along this line, offering public services is often assumed to benefit citizens. Digitalization of government directory, forms, tax filing, vehicle registration, etc. makes such traditional services convenient for citizens. The flourishing of #GovTech projects all over the world demonstrates the significance of such e-Government initiatives in transforming the infrastructure of public administration and governance. Although benefiting citizens is the common starting point, not all of them stress the importance of citizen engagement.

This way of defining civic tech tends to **treat citizens as beneficiaries, or customers who receive benefits from using these digital services, although with complaints**. For instance, a paper highlights the problem of treating citizens as complaining customers when citizen hotlines are set at the US Congress as customer service lines [52]. We can see that this way differs from other definitions that factor in citizens' own input on what tech is good for them and how to design the tech. The second layer of definitions emphasizes "by citizens", not merely "consumption".

*3.2.2 By Citizens.* This layer of definitions **treats citizens as one category of civic actors, alongside or relatively independent of other actors such as the government and the market forces.** As Schuler defines cooperative intelligence, he contrasts civic aspirations against the logic of political and commercial winning.

"Prefixing the modifier 'civic' to 'intelligence' signifies that it is something that is activated in service of civic aspirations. The term acknowledges the potential of an intelligence that can be cooperative, that isn't evaluated or accomplished by "winning" or by profits or market share." [77].

The conceptual distinction between citizens, governments and commercial entities is an important one. If we only consider "for citizens", the boundary among the three types of actors becomes blurred. Some e-Governance projects take advantage of the market forces (e.g., vendors) to develop digital services, with minimum input from citizens. For instance, in one such definition below, civic tech includes clearly commercial applications.

"(c)ivic tech is a term that refers to the diverse ways in which people are using technology to influence change in society. The breadth of civic technologies is wide and comprises a large pool of technologies for i) governance (e.g., MySociety, SeeClickFix), ii) collaborative consumption (e.g. Airbnb, TaskRabbit), iii) community action (e.g. citizen investor, GeekCorps), iv) civic media (e.g. Wikipedia, Global Voices) and v) community organizing (e.g. WhatsApp groups)". [47].

However, if we emphasize the component of "by citizens", we have to first identify who citizens are, not just vaguely "people". Ironically, clear definitions of citizens are the rarest in our search. From our review, we found that ordinary people (the young, the old and the minority), activists, community members, and non-governmental organizations have all been part of "citizens". As long as the actors are not part of the government or acting to make profits, they are understood as citizens.

After identifying who citizens are, "by citizens" assumes some level of citizen engagement. We observed two engagement approaches in the literature: The interaction approach focuses on the interrelationships among multiple actors, especially the government. The concept of "digital civics" follows this approach.

"Digital civics seeks to design systems based on developing and brokering relations between government and citizen." [12].

"digital civics...seeks to design tools that support relational civic interactions across multiple categories of civic actors" [6].

The interaction is further defined as partnership or coproduction, such as in this definition of "we-government":

"the re-emergence of citizen coproduction—whereby citizens perform the role of partner rather than customer in the delivery of public services" [65].

The empowerment approach puts more emphasis on highlighting the autonomy and agency of civic actors, such as understanding their priority, building their competency, and facilitating their actions. For example, civic learning engages citizens in learning activities to enhance their competency.

"Civic learning...supplies the learner with the knowledge, skills and values they need to be citizens who actively participate in their local communities and take responsibility for improving and understanding them." [69].

*3.2.3 Democratic engagement.* **Not all engagement by citizens is democratic**. This layer of definition treats citizens as participants who have to be engaged in democratic manners. One definition emphasizes that civic tech "facilitates democratic governance among citizens" [73]. "Democracy is an approach to governance in which people meaningfully and intentionally participate in the decisions that affect them and other members of the group." [78]. Engagement in decision-making by citizens characterizes democracy, although the engagement modes vary according to the democratic models. In [59], democratic engagement is classified based on three ideals: representative democracy that focuses on voting for delegates; referendum democracy that focuses on voting for decisions; assembly or deliberative democracy that focuses on discussing before reaching the decisions. Each model needs legal and bureaucratic frameworks in place in order to work properly, and the absence of such frameworks makes such democratic engagement unavailable or incomplete. Therefore, democratic engagement by citizens is the most stringent definition of civic tech, considering that many countries do not operate under fully democratic systems.

## 3.3  Definitions of "Tech"

Compared to the multi-layer and mixed definitions of the civic, most of our reviewed works take technology for granted. In short, ICTs occupy the central stage.

"Of particular interest is the role of information and communications technologies in shaping contemporary engagement and actions, as referred to as 'civic media' and 'digital civics'" [54].

The Knight Foundation report [25] identified further the types of ICTs that overlap with civic tech purposes during 2000-2012, including crowdfunding, P2P sharing, open data, data utility, data visualization and mapping, community platforms, feedback tools, public decision making and voting tools.

There is a rising trend that probably learns the lessons from prior (failed) efforts of inventing tools only. More studies have started to call for offering design solutions, building processes, and transforming infrastructure. One such expanded "tech" is hackathons [39]. As Pilemalm puts it, the civic tech we are building:

"... often need to deal equally (or more) with organizational/institutional transformation than exclusively ICT development..." [65].

Putting civic and tech together, we can roughly summarize that Civic Tech refers to **the usage of information and communication technologies to benefit citizens.** Some of the solutions include engagement by citizens and even fewer engage citizens in democratic manners.

## 4  THEORIES IN CIVIC TECH

We searched for mentions of theories in the 224 articles. Theories across a broad range of disciplines in political science, social science, technology, and design science have been used to explain civic tech and the contexts surrounding its usage. Some theories seek to describe the political environment in which civic tech is built and the social structures that underlie it. Some seek to describe the users and applications of civic tech and explain the interactions that occur between people and technology. Some seek to describe the goals and creation processes of civic tech. The diversity and variety of theories provide a rich understanding on the various dimensions of civic tech. Through a survey of the review corpus, these theories have been identified under two broad categories: civic vs. tech theories. Each category of theories is briefly described in Supplementary Materials Table S1 and S2.

### 4.1 Civic Theories

**Democracy-related theories describe the types of democracies that have been conceived and practiced in democratic countries** where civic tech is largely produced. Saward puts forward three forms of democracy: 'representative' where decisions are made by elected authorities, 'referendum' where citizens' votes are taken on the matter at hand, and 'assembly' where action is taken based on the result of citizens' discussion, with the latter two falling under direct democracy [74]. Parallels of Saward's model can be drawn to Van Dijk's models of democracy that further differentiates the types of democracies according to its goals and means, and references the role of ICTs in each of them [17]. In particular, representative democracy corresponds to the competitive model where computerized information campaigns are used, referendum democracy corresponds to the plebiscitary model where telepolls are used, and assembly democracy corresponds to the libertarian model where online forums are used. The prevalent use of ICTs to aid in governance and democratic processes has also been termed 'e-government' and 'e-democracy' to surface the increasingly intimate and crucial role of ICTs in democracy. Within the scope of citizen participation, 'deliberative democracy' and 'agonistic pluralism' have also been offered as two contending modes of citizen discussion where the former aims towards reaching consensus in decision-making whilst the latter advocates for competitive dissensus.

**Citizen-related theories are concerned with civic-mindedness and the level of engagement people have towards civic matters**. Different people have varying levels of interest and involvement in political issues. Those unconcerned may decline to vote in elections whereas those with keen interest may go on to become activists. Through increasing

participation and engagement of citizens in civic matters, civic intelligence and social capital are generated. Collective knowledge from the informing and self-discovery of citizens contributes towards collective awareness and action for the common good such as identified in the United Nations' Sustainable Development Goals[5]. Interactions that lead to better understanding and bond-building among people also paves the way to a more concerned, caring, and cohesive society.

**4.2 Tech Theories**

**The set of behavior theories seeks to explain human behavior and the interactions that occur within groups of people**, often serving as a foundation to understand the interaction between people and technology. Broadly, the social theories cover motivation, interactions within social structures and ethical considerations that are present in all human activities. **Technology-Use theories zoom in on the relationship between people and technology through the use and usage of technology**, and the impacts brought about by technology. These theories provide an intricate understanding of the subtleties of human behavior from individuals to societies and the ways they have been shaped by the rapid advancement and adoption of technology.

**Design theories provide various ways of thinking about and approaching the development of technology**. One set of design theories looks into the ideals that technology should aim towards. These theories take a critical perspective into who uses the technology, what the technology is used for, how the technology has been used, and more importantly, the converse situation. Doing so, the theories surface important threads of consideration that can be incorporated into the design of technology for more humanistic technological development. The other set of design theories focuses on the implementation of technological development. These theories describe the processes taken to design the technology such as by establishing a strong understanding of the users and the situations served, and by examining the potential impacts of technology through extensive investigation and synthesis.

Depending on the goals, civic tech studies can consult these theoretical frameworks to construct a solid foundation for tech development be it in the design, prototyping or evaluation phases. Democracy and citizen-related theories can inform the participation mechanism that the civic tech supports. Behavior theories like self-determination theory and flow theory can supplement civic tech design that seeks to bolster user interest and engagement and is sustainable over time. Design theories can inform the approach in which tech development takes. And technology-use theories like the technology acceptance model can be effective for evaluation. These theories have been built upon fundamental knowledge in social and behavioral science and can be more greatly utilized to enhance civic tech development in concern with people and society. In civic tech development, it is also crucial to mindfully embed design theories in the social contexts that the technology is used for. Society is a complex system and every silo has its peculiarities. To nudge or even engineer complex social factors, a good understanding of user psychology will be of great support. As civic tech is concerned with supporting social interactions for democratic processes, having more integrated theorization across the socio-technological domain that pulls together knowledge from both fields will be crucial for the future development of civic tech.

---

[5] https://www.un.org/sustainabledevelopment/sustainable-development-goals/

# 5 HISTORICAL DESCRIPTION OF CIVIC TECH RESEARCH

## 5.1 Publication Venues

In total, the 224 articles in our review corpus were published in 62 venues (Figure 3), where the Proceedings of the ACM on Human Computer Interaction 'PACM HCI' is subsumed under the ACM Conference on Computer-Supported Cooperative Work and Social Computing 'CSCW' due to a change in publication format since 2018. The articles were published over a period of 21 years, with an average of 3.61 articles per year. The top five publishing venues are CHI (n=54), CSCW (n=41), dg.o (n=15), DIS (n=11) and ICEGOV (n=10). The remaining 57 venues published an average of 1.65 articles per venue, with a standard deviation of 1.40. The earliest article was published in JCDL 2001 that discussed the use of community information systems to supplement information needs for daily problem solving [64]. Thereafter, the number of publications increased in an exponential fashion up to 2020, indicating a gaining momentum in the interest of the field. Most notable is the marked presence of review papers and critical essays (e.g., [59,78]) in 2020, signifying that there is consensus that the field has reached a certain stage of maturity, making this review a timely one.

The recent trends show that in 2012, there was a significant jump in the number of papers from 3 to 12. Since then, the numbers have kept at double digits. Consistent with what we introduced in the related work section, the overlaps between civic tech and e-government fields existed but were not large. ICEGOV and dg.o published 1-3 papers every year since 2012, and had no papers in 2019. Another observation is that there were quite a few papers, almost half in 2015, 2018, 2020, which appeared in venues other than CHI and CSCW. This means that the two most prestigious ACM venues were not fully open to the civic tech scholarship, probably due to its early adult status.

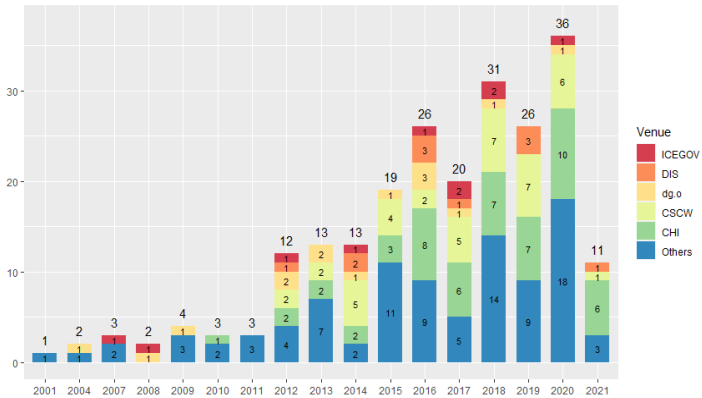

Figure 3. Number of publications in civic tech by year and by venue. 'Others' include the remaining 57 venues (e.g., C&T, NordiCHI, OzCHI, …).

The top five venues are concerned with human-computer interaction, communication, and governance, fields that are central to civic tech. Yet, there is great potential for several other venues that are also strongly relevant yet have only seen a low interest in civic tech to critically contribute through their various areas of expertise. For example, with a focus on software engineering in society, the ICSE-SEIS track (n=1) could lead the discussion on the technical design and engineering of civic tech. GROUP (n=2) may provide insights on how small groups to large communities interact with each other through civic tech. WWW (n=2) may examine the effectiveness of social media on political participation and the negative externalities of misinformation. And last but not the least, PDC (n=2) may push for efforts in incorporating participatory processes that involve citizens in the design and use of civic tech.

**5.2    Research Methods**

An overview of the research methodologies employed in the review corpus is shown in Figure 4. There are 17 articles without a clear or no research methodology, indicated as 'N/A'. These are largely theoretical research papers that seek to present new ideas based on the discussion of existing concepts or situations (e.g., [77,78]). The remaining 207 articles are analytical research papers and review papers with the following breakdown of research methodologies: quantitative (n=40), qualitative (n=113), and mixed methods (n=54) where both quantitative and qualitative methodologies are used.

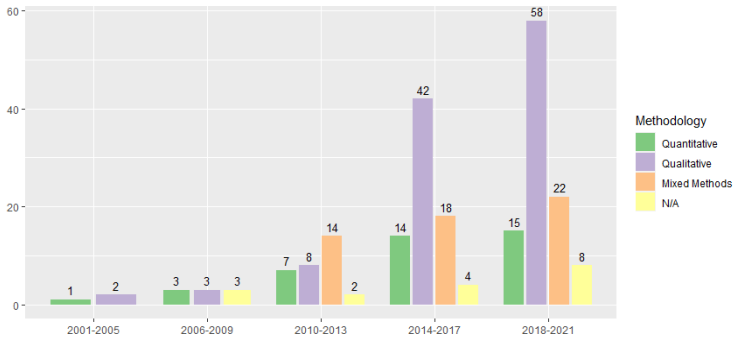

Figure 4. Research methodologies used in civic tech publications.

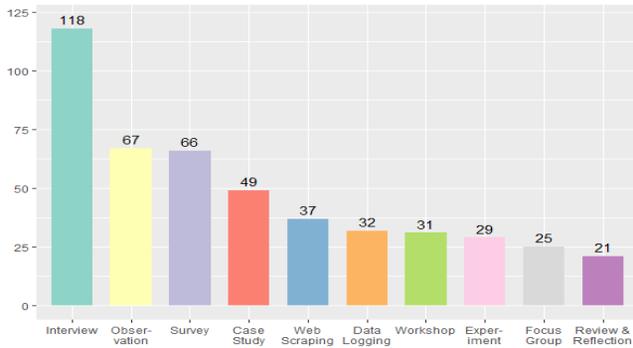

Figure 5. Research methods by research methodology of civic tech publications.

Throughout the years, there have been significantly more qualitative studies on civic tech (see Figure 4). This is due in part to the people-centric focus of civic tech whereby the understanding of how people use technology for various civic activities and how they perceive those experiences have become of great interest to researchers (e.g., [66,68]). Another factor is the interest in examining how specific communities communicate and collaborate through the use and appropriation of technologies (e.g., [19,24]). From Figure 5, interview, observation, survey and workshop are the more popular research methods for qualitative studies. The usage of workshop as a method might be a unique tradition in the civic tech field, compared to other ICT development. This method follows theoretical ideas such as participatory design and resonates with hackathons that are common in the practitioner arena.

Contrastingly, quantitative studies saw a rise in interest in 2010-2013 and have been consistent over the later years (see Figure 4). In 2010-2013, there was a marked number of articles on big data analysis, particularly regarding interactions and behavior on social media (e.g., [30,57]). This could be related to the rising popularity of social media and the big data these platforms collected from users. From 2014-2021, there is a greater diversity of articles covering big data analysis and practical applications like predictive modelling and machine learning techniques to address issues related to social media (e.g., [11,58,67]). Earlier quantitative techniques used descriptive and inferential statistics in surveys and experiments while later more complex analysis was conducted on larger datasets using big data analysis, modelling, machine learning and natural language processing. From Figure 5, web scraping, data logging and experiment are more greatly used in quantitative studies.

A combination of the two, mixed method studies incorporate the greatest range of research methods. Many of these studies (e.g., [35,70]) are interested in understanding how people perceive and use technology. For example, interviews and surveys are used to solicit user perceptions and feedback, and data logging is employed to capture usage behavior on the technology. Such studies then provide a more holistic view on the interactions between people and civic tech.

**5.3 Discussion of Historical Findings**

Looking at the 20 years history of civic tech research, the earlier studies started from describing and understanding the use of ICTs by civic actors such as NGOs [71,82], activists [5,40], volunteers [69,83], and governments [16,35]. Youths, as a group of citizens who are disinterested and inactive in participating in traditional civic actions, have consistently been a target user group [9,45,48]. These descriptive findings often inform designers about what is needed to be designed and how the design can be effective.

As the field develops, later studies are more likely to go beyond pure descriptions and offer solutions, either in the format of a framework/process or a piece of technological tool or a mixture of both. The types of civic technologies developed and used correspond to the technological trends in the larger field of ICTs. For example, as social media data become available and Natural Language Processing tools mature, machine learning starts to be applied to analyze civic content on social media [2,88,90].

Moreover, civic technologies echo the changes in social and political climates. For instance, news and journalism technologies were common till 2015 [10,20,56], with several projects trying to frame social media content from ordinary users as citizen journalism or civic media creation. However, starting from 2016 when American President Donald Trump popularized the term "fake news", this line of civic tech research almost disappeared and was replaced by studies that develop Machine Learning classifiers to detect fake news [48,67]. Another example is that when many liberal democratic countries legalized open government data, data analytics tools were built to take advantage of such data [11,33,41,43].

The development of civic technologies cannot be separated from the evolution of the technologist community and its subcultures. The hacker and maker movement provided technologists who are attracted to developing non-commercial and open tools a commune space to connect at a global scale [37,47]. The world-wide locations of hackerspaces set the precedents for technologists to find innovative ideas and similar minds. The maker movement, following the philosophy of the open-source movement, aimed to open the blackbox of technologies to the public. These existing practices set the foundations for civic tech groups such as Code for America to emerge in 2012, the year in which civic tech publications had a first jump. Civic tech inherited both the spirits and practices from its forerunners. Hackathons, for example, started as gatherings for technologists to come together and solve problems in collaboration. Now hackathons have become a common practice widely used by governments, organizations, educators, and corporations. Different from them, civic technologists treat hackathons as a method to engage citizens [39].

The usage of technologies evolved from either an existing platform such as Twitter or a tailor-made technology made by the authors, to a range of combinations. Some combinations introduce small nudges to existing platforms [55], and some completely reinvent existing platforms to suit their needs [62]. The general tone towards civic tech has changed from overall positive to cautious. A contrast was seen between earlier papers that discussed how new media can help civic actors [63] and later ones that focused on how risky and even dangerous popular social media might be to civic actors [21].

## 6   ANALYTICAL FRAMEWORK AND KEY INSIGHTS

Drawing upon the theoretical review and discussion, we propose a Civic Tech Framework that serves as a transition step to connect conceptual thinking with design and practical thinking. The framework can be summarized in one sentence: **who used what tools through what process in what contexts**. While most conceptual discussions understand "who" in terms of their civic status, such as ordinary citizens, activists, NGOs, government, etc., this framework proposes a set of roles to understand the actual action makers in civic tech. The focus shifts to find out who fund, design, implement and maintain the piece of civic tech, and who are the partners and participants. The second component of this framework tries to sort through digital tools that range from modifying existing tools to inventing tailor-made tools. Process is singled out as a critical component that highlights solutions that are more than just ICT tools, which are visible in not only the design but also the evaluation and implementation processes. Participation mechanism that is actually used to engage citizens receives special attention here. The last component of this framework is to detail the contexts in which civic tech is used, as different issues often involve different problem spaces and different countries prioritize different issues. The framework is presented in a circular format in order to emphasize the mutual influences among the factors that evolve over time. For example, as the design process goes, partners might be introduced into the project; or as new funders join the project, they may suggest changing the issue context.

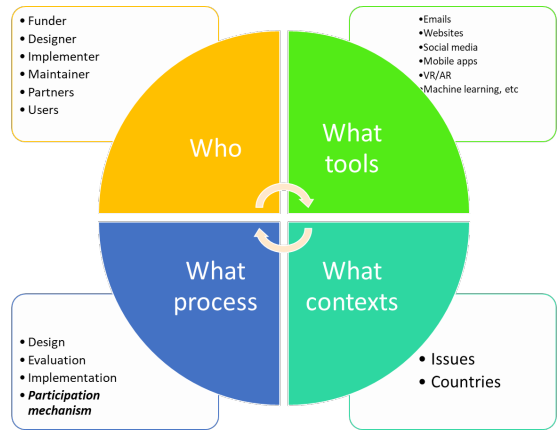

Figure 6. Civic Tech Framework: A circular framework describing civic tech developments.

Using this framework as a guide, we discuss what the field as a whole has learned from existing practices by focusing on the four key insights: 1) Who have been serving the roles in civic tech projects? 2) What technologies have been designed and/or used? 3) What processes have been followed, and how were citizens engaged? 4) In what contexts have civic tech initiatives been built and applied?

## 6.1 Who Are Serving the Roles?

Building technology always requires resources and building civic tech faces special challenges because profit-making is not the primary goal. Investing resources becomes difficult when no profit returns are expected. Among various actors, the government is apparently more resourceful compared to others. Our analysis (see Figure 7) shows that the biggest funder of Civic Tech is governments, leading other types of funders with a large margin. Although governments fund civic tech, it is intriguing to see that governments are not a regular partner in the design and implementation of such tech. Both foundations and universities/schools support civic tech projects as the second-tier funders. The smallest number of funders falls into the category of commercial companies, which often support civic tech as part of their corporate responsibility program. Universities/schools and companies not only fund but also partner in civic tech projects, showing a balance in the roles. Organizations are more likely to play the partners role than the funder role. While local communities are not able to provide funding, they sometimes play the role of partners in civic tech.

When looking at the process of civic tech, designers play the role of coming up with the tools, coordinating the process, and sometimes actualizing the usage. Users are the ones who will use these technologies. Our analysis (see Figure 8) finds that academics take the lead in playing the role of designers, followed by companies because many tools are adapted from existing commercial products such as social media. Government, organization, community, and citizens all have low presence in the designer role. In contrast, they often play the user role: both general and specific citizens (e.g., students, youth, the elderly, and people with disability) are the majority of the imagined users of the civic tech being designed. Organizations and communities frequently participate in the process as users but there are relatively few instances in which government officials are the study users -- a surprising finding if we consider how much civic tech is funded by the government.

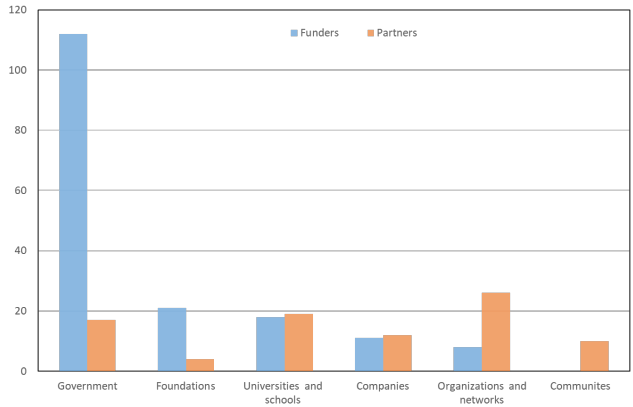

Figure 7. Funders and partners in civic tech projects.

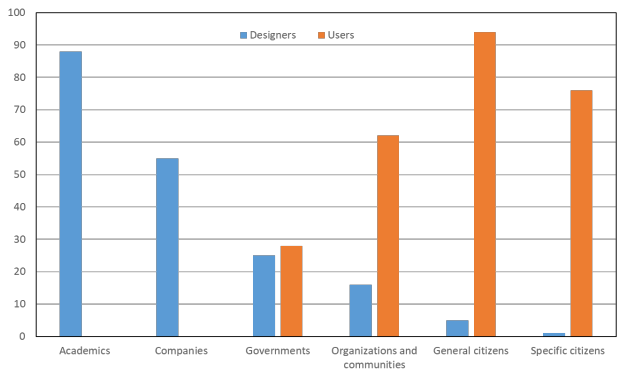

Figure 8. Repartition of the designers and users in civic tech projects.

When we link actors with different civic status to the functional roles they play, we observe a pattern of division. Stakeholders such as funders and partners often differ from designers and users. Academics play a key role in making civic tech happen but have to depend on other entities to provide resources. Although the tech is meant to serve civic actors, these actors are rarely involved in design and mostly participate as end users. Due to the division, tensions between the actors serving different roles exist. The first tension is seen in governments vs. end users. For example, a study found that when the grassroots desire to innovate, "local governments' rigid compliance with statutory obligation" prevents new solutions to be used [18]. In another case, the state government transfers the provision of social services to NGOs, through providing financial and other support. but the joint actions still need a lot more mechanisms such as legal, institutional, and communication ones to be effective [82]. Governments need to go beyond the role of grant providers and level up their engagement, such as building infrastructure, fostering civic tech communities, and co-designing and implementing the solutions.

Another tension exists between the civic tech projects' short time span and the long-term needs of communities and citizens. Sustainability has been a recurring challenge many have discussed. Civic hacking projects, or hackathons, have been used in many civic tech initiatives. However, "the prototypes were rarely implemented, and hackathon participants had no time for 'real footwork' to build coalitions and trust with partners and citizens." [39]. Our findings show that not only hackathons, but also actual technologies built are not well sustained: about 40% of the civic tech tools are no longer in use. Among the ones that are still in use, a majority of them (91%) are existing platforms that are often commercial applications. Civic tech tools need to take advantage of "off-the-shelf technologies" or existing tools and put more effort into process design in order to achieve better sustainability.

### 6.2 Which tools are used?

Looking at the technology used for Civic Tech projects shows interesting trends in the field. As shown in Figure 9, existing works in the field tend to mostly rely on **web-based platforms** (n=18), **physical devices** (n=13) and **mobile apps** (n=9). The use of mobile apps and web-based platforms [62] allows researchers to expand the spatial scope of their research by deploying potentially on the state-, province- or country-level. The number of Internet users has quickly increased from 1.4 billion in 2008[6] to more than 5 billion in 2020. Many early works with web-based platforms were thus done in countries with high rates of Internet users, e.g. USA [42] or Finland [31]. In parallel, after the iPhone was released in 2007, the number of smartphone users increased globally from 1.06 billion in 2012[7] to 3.6 billion in 2020. The prevalence of smartphones is changing Internet usage habits as more than 52% of the global data traffic goes on mobile devices in 2020[8]. With technology becoming more widespread, civic tech projects should be able to leverage either web-based platforms or mobile apps to reach large numbers of citizens. However, web-based platforms require devices that may display the platform in an optimal way which may exclude citizens who do not own a tablet or desktop computer. Mobile apps may exclude users who do not own a smartphone compatible with the app, as iPhones cannot run Android apps and vice-versa.

To reduce the issue of access to technology in a local community, some researchers use physical devices that are deployed at specific locations within specific communities, making any passerby a potential user of the technology [85,91]. Interaction with such systems tends to be quick and simple (e.g., answering a question with a few preset choices [85]). Physical devices allow local people to vote (or choose options), share concerns or give their opinions on community life. This type of tool only works on a small scale but does not prevent anyone from participating as these systems usually offer simple interactions to catch attention, making them noticeable by the passer-by.

Since 2012, another trend emerged: **dataset papers** (n=30), works in which public data were scraped from specific internet platforms, e.g., Twitter [56,57] or Facebook [34]. As the number of social networks users is also growing rapidly (2.7 billion Facebook users in 2020[9]), running civic tech projects on dedicated Facebook pages, or scraping data from the site, allows researchers to reach a large number of potential users easily, with less reliance on tailor-made platforms.

---

[6] https://www.internetworldstats.com/emarketing.htm
[7] https://www.statista.com/statistics/330695/number-of-smartphone-users-worldwide/
[8] https://www.broadbandsearch.net/blog/mobile-desktop-internet-usage-statistics
[9] https://www.statista.com/statistics/264810/number-of-monthly-active-facebook-users-worldwide/

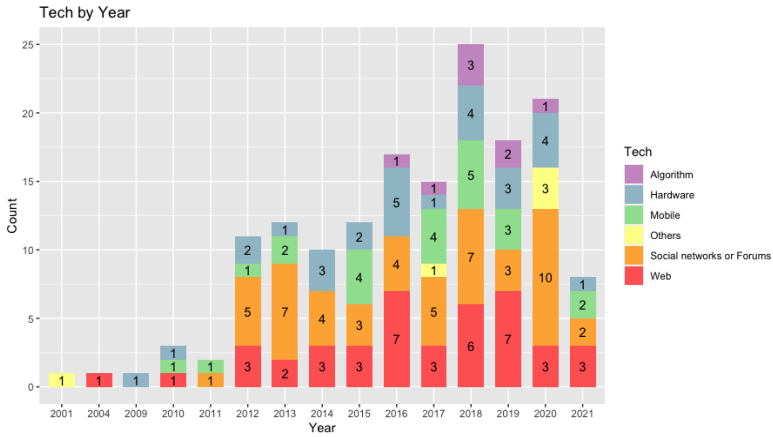

Figure 9. Type of technologies used over the years. Papers relying on surveys or interviews are not shown. Algorithms suggests that the contribution of the paper is on how the data is processed, or created metrics. Others include Virtual Reality (1) and UI Prototype (1).

In most cases, the tools are developed for the sole purpose of the research project (i.e., tailor-made tools). As such, development for such technologies may take months and they may simply not be reusable in other contexts. There are also other factors explaining the downfall of tailor-made tools which comes from the technology, languages and libraries used. Software may become obsolete within a few years, reducing the ability of researchers and practitioners to find suitable developers to tweak existing platforms. Looking at web-based platforms specifically, Ganoe et al. [26] used Java Servlets, while Klein's MIT Deliberatorium [45] was developed in Lisp: two technologies that have disappeared except for niche applications. Even PHP, which used to be one of the most popular languages in the early 2010s and was used by Mahyar et al. in 2018 [51] and Perrault et al. in 2019 [62] is slowly being phased out. Similarly, CONSUL[10], developed in Ruby, uses deprecated packages and cannot be deployed on recent popular Linux distributions (2018 and above). All these limitations may explain the interest in getting data from existing sources (e.g., dataset papers) which are less dependent on changes in technology, as long as the main source of data (e.g., Facebook) maintains an API to access the data. In contrast, tailor-made tech tools (starting 2018) require time to develop, test and improve and tend to be usable to produce one or two papers at a time.

However, ready-made technologies, especially the highly commercialized ones, have their own challenges. Although their sustainability and availability seem to be better than tailor-made technologies, the critical issues associated with technologies that serve profit-seeking are many. For example, social media tools collect a large range of user data and these private data may be used for unethical reasons (e.g., Cambridge Analytica). The security and safety of these data are in question when platforms are hacked and private data are leaked. Moreover, driven by commercial interests, these social media tools on the one hand use algorithm to manipulate user behaviors, one the other hand do little or nothing to correct misinformation or reduce discrimination. All these risks and limitations of such tools point to the difficulty of choosing or designing the right civic tech tools.

---

[10] https://consulproject.org/

In summary, civic tech research benefits from both tailor-made tools and existing technologies, with the latter bringing breadth as it may last longer but may be constrained by their commercial nature, while the former bringing depth as researchers may carefully design the tools to suit civic purposes. The tailor-made tools can increase citizens' problem awareness, critical thinking, common ground, civic learning, and consensus building. The pressing problem is their sustainability and availability.

### 6.3  What participation mechanisms are there?

In the creation of civic tech, users play a variety of roles that can be surfaced through the diverse set of design processes used (Figure 10). In user-centered design (n=3) and typically iterative design (n=10), the emphasis is on understanding user requirements and developing tech to address them (e.g., [29,70,86]). In the early stages, users are involved in extensive surveys and interviews for the identification and prioritization of user needs. Thereafter, users are recruited to evaluate the effectiveness of the tech and to provide feedback for improvement. In this way, the user can be seen as a patient whose symptoms are to be consulted and diagnosed, where ultimately the doctor, who is the expert, prescribes the cure. While there may be multiple iterations throughout the process, interaction is generally one-way, from the user to the expert.

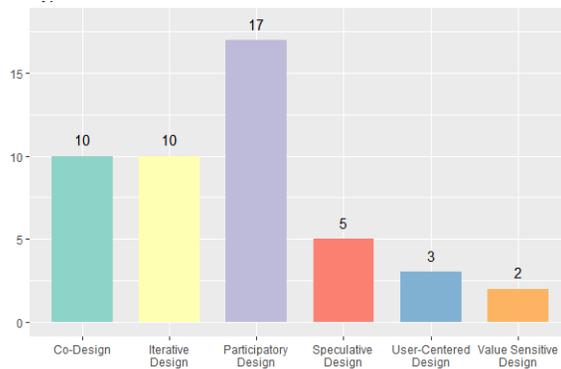

Figure 10. Design processes used in civic tech development. Frequencies are only counted for articles that clearly state the design process.

In contrast, users play a highly prominent role when participatory design (n=17) and co-design (n=10) is employed for the creation of civic tech. From the outset, users and other stakeholders are recruited to actively engage in the co-creation of civic tech from the early discovery phase to the later ideation and evaluation phases (e.g., [20,65,91]). Beyond surveys and interviews, workshops and other forms of long-term engagements are conducted to provide opportunities to gather the various users to lead in the discussion about the problems they face and to offer suggestions, implement and test the solutions they have designed for themselves. Interaction between the users and the experts occur two-ways through mutual and constant collaboration where users are regarded as dynamic, creative, and enriching partners whose first-hand experiences and insights into civic issues are especially valuable.

In some cases, there is a less defined division of roles where users and experts may either operate independently or collaborate. In speculative design (n=5), designers can be the users, experts, or both. When users are the designers, experts take a less prominent role such as moderating the speculative design activities. When experts are the designers, users take an evaluative role, much like in user-centered design. In the last case, the dynamics is more like in participatory design. This difference is because the emphasis of speculative design is on designing the future, where the choice of the designer becomes a matter of intention instead.

Finally, instead of direct participation, users become a presence for examination as part of a wider set of entities. In value sensitive design (n=2), the interest lies in understanding and incorporating human values in the development of technology. Designers consider the values of diverse stakeholders, that includes users, and focus on studying the consequences of the technology on society and the world when developing them.

### 6.4  In what contexts?

Contexts can be understood as a component that includes at least two dimensions: issues and countries. We define issues as civic problems to be solved or civic challenges to be addressed. Our review shows that specific issues received varying degrees of attention in civic tech studies. The range of issues we found is large, covering everyday livelihood to emergencies and disasters (see Figure 11). The most frequent issue is to address challenges concerning democracy such as (the lack of) citizen participation in political decision making and engagement such as community building. Public service and governance issues are the second highest, including providing services to the public (e.g., food, housing, transportation) and involving the public in policy making (e.g., policy consultation). The third highest issue is to take advantage of new media to disseminate useful information and fight fake news. The categories of urban planning and neighborhood/local issues help geographically defined communities to address various local matters including public displays, arts installations, feedback and consultation tools, and collaboration events. Specific issue areas focus on education, accessibility and minority groups, sustainability (mostly climate change issues) and environment, crime and safety, and disaster. We can see that the dominant issues are those closely connected to the general solutions to amplify citizens' voices and actions in collective decision-making. Sporadic events such as disasters and crimes are paid least attention to. However, the relatively low focus on sustainability and environment, accessibility and minority, and education are worrying, given how climate change has become a globally urgent issue and how social movements such as #BlackLivesMatter has shaken established social hierarchy - a finding that suggests future efforts to be put in addressing such issues.

Among studies that have specified the locations, we find that the wealthiest parts of the world lead the civic tech developments with a wide gap. As shown in Figure 12, a vast majority of the studies are located in the US, accounting for 42% of the studies. UK and Europe follow closely and account for a combined 35% of studies. The Asia Pacific region including Australia and New Zealand contribute 13% of the studies. The rest of world (including non-USA American countries, Africa, etc.) in total contribute 5% of the studies. Country-wise, US and UK are the dominant players in the field, with not only developing countries but also other developed countries lagging far behind. In other words, the geographical imbalance is less of a developing vs. developed country divide, and more of a US+UK vs. the rest of world divide. Moreover, we find that some studies (5%) take advantage of the virtual space that goes beyond national borders to locate their research efforts, a direction the field may pursue in order to address the severe geographical imbalance. Different countries often face different challenges. Issues such as electoral fraud [80], corruption [30], and the solutions to address such problems (e.g., civic reporters [56]) were more often explored in studies from the Global South.

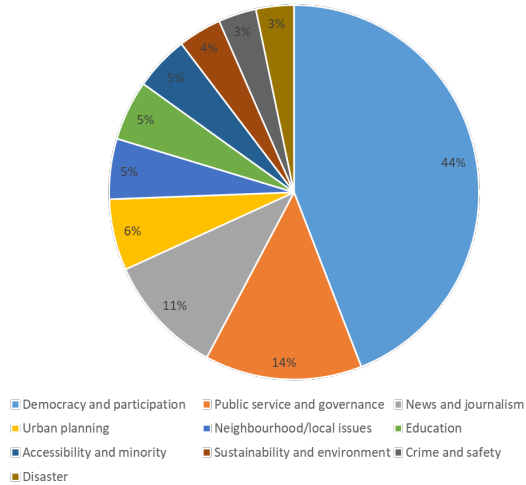

Figure 11. Issues addressed in civic tech projects.

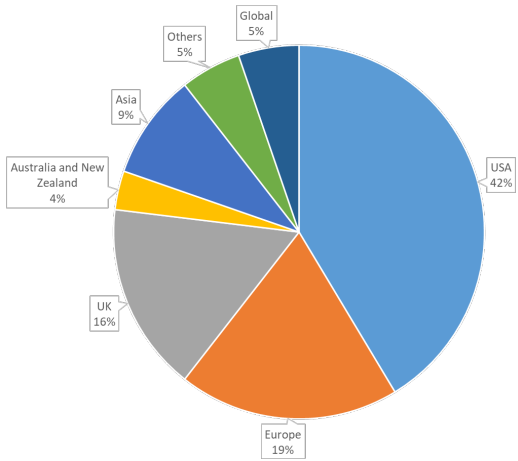

Figure 12. Study locations for civic tech projects.

## 7  DISCUSSION AND RECOMMENDATIONS

Our review has revealed that the Civic Tech field has accumulated knowledge regarding both civic and tech aspects. Moreover, the field has evolved along with technological developments to take advantage of the latest technological tools. The most important achievement is that a field that pays equal attention to research and practice and receives support from various stakeholders has come into shape. In the discussion, we highlight several shortfalls related to civic tech, hoping to join the collective effort of further developing the field. These shortfalls come under two categories: that civic technological development is yet to serve civic purposes well and that mainstream technological development has ignored or even hurt civic values. The former describes the limited outreach and effectiveness of otherwise successful civic tech, while the latter describes the areas of negligence and failure in civic tech.

**7.1 Why did civic technological development not serve civic purposes that well?**

*7.1.1 Limited availability and accessibility.* The lack of adequate technical infrastructure as basic as the internet service [7,88] fundamentally prohibits the development and operation of civic tech. Correspondingly, a low level of digital literacy hinders the learning and adoption of civic tech [12,16]. And finally, even given that the first two factors were not a concern, a low awareness of available civic tech resources, due in part to vastly more popular social media technologies that serve as a barrier to entry for new technologies of public engagement, dampens the level of participation [27]. Together, these factors inhibit the availability and accessibility of civic tech and are more prevalent in less-developed areas and countries.

*7.1.2 Low trust.* With data breaches continually making headlines over recent years, privacy and trust remain a key issue hindering the adoption of civic tech. This is especially so for government-led systems where it is unclear how highly sensitive data is used and for what purposes [12,36]. These issues are also found in other public systems. There are concerns on anonymity in deliberative civic tech particularly within political contexts where there tends to be fear of expression and association [72]. Moderation has also been shown to cause strong resistance towards participation in the deliberative process [62]. These signify that civic tech researchers should not only be concerned about a lack of trust towards the authorities and administrators of the systems, but also among citizens themselves.

*7.1.3 Inadequacy of technological solutions for social issues.* Another issue is the inadequacy of technological solutions in approaching the intricacies of social behavior. Social work is a prime example where the developmental, preventive, or remedial work is of a highly sensitive and tailored nature and must be dealt with using utmost care and caution. In this way, technological interfaces are unable to replace the intimate relations that front-line social service professionals must establish with the people they serve and may at-best only play a supporting role [9]. The use of technology such as simulation games and virtual reality to develop empathy must also be cautioned as the immersive context of these solutions may instead lead to a detachment with real circumstances, thus running counter to the objective [61].

*7.1.4 Loss in sustainability.* For civic tech where prolonged engagement is desirable, it is essential to address loss in interest over time. This requires a balancing of the usability and complexity of civic tech systems. Unchanging forms of interactivity and content may be easy to learn but will quickly fall into the mundane. On the other hand, too many features in a system increases the difficulty of use and may even obscure the user from the true purpose of the civic tech [83]. While it is hard to keep users interested, it is equally challenging to maintain resources to sustain the civic tech tools built [35]. Without formally integrating such tools into the political system, many brilliant civic tech inventions have waded after the resource pool was drained.

**7.2 Why did mainstream technological development ignore or even hurt civic values?**

*7.2.1 Neglecting marginalized groups.* Parallel to civic tech development, mainstream technological development can come to ignore civic values. Public technology has traditionally been designed for the general public, thereby neglecting marginalized groups of people who may require added functionalities such as accessibility settings or translation to mother languages [8,68]. For civic tech serving all citizens, the marginalized groups are often of an equal if not greater concern and if they are excluded from the civic processes on a fundamental technical level, this limits the accessibility and functionality of the tech.

*7.2.2 Underestimation of the complexity of civic issues.* Most mainstream technological developments do not take the civic as their primary design goal. Cautions have been made for this naïve thinking that tends to reduce the social and political aspects of civic issues to the purely technical that can be fixed by simply applying the right technological solution [81]. Given the complexity of civic issues, there is just no one approach that can adequately address all problems, and what is needed is instead a variety of both social and technical solutions [78]. For example, Schuler raises non-democracies as an entity for consideration [78]. Tech for these countries and collectivities would require new models of participation that have yet been explored in the literature that leans towards democracies.

*7.2.3 Technological oversights.* In other instances, some mainstream technological development stands the possibility of hurting civic values. With the rise of artificial intelligence in recent years, the presence of biases in the algorithms has become widely raised, leading to ethical concerns on the legitimacy and transparency of such systems [14,23,35,53]. Another aspect is that with big data, data-driven approaches are increasingly preferred by governments to analyze and address local issues, further stamping out opportunities for direct engagement with citizens and undermining community voices [16,52].

*7.2.4 Harmful use of tech.* The mainstream technological development also suffers from the harmful and exploitative use of tech. While not meant as a civic tech, social media and online forums have undeniably become one of the greatest tools of political participation. As much as they have enabled an unprecedented scale of engagement, the ease of access has also empowered bad actors to spread falsehoods and prejudices to an extensive global network, often inciting online incivilities and offline violence that constantly put the civic under threat [53,72,78,90].

**7.3 Recommendations**

Looking ahead, there are many areas where future civic tech endeavors may be advanced. These areas are not necessarily straightforward to address but are vital in paving the way towards more holistic and effective civic efforts.

*7.3.1 Educate citizens.* Informing citizens on the state of national affairs and equipping them with the skills to understand data has been found to contribute to more constructive deliberation as users are able to make better sense of the information they have [53,62]. With low civic and digital literacy remaining as inherent barriers to civic engagement, sustained efforts on the part of the government, education system, guardians, and individuals are required to develop these skills [78].

*7.3.2 Improve trust relations between stakeholders.* Given the sensitive nature of affairs that certain civic techs manage, there have been calls for greater transparency and accountability of these systems, particularly for those led by governments [16,66]. Trust between the users and owners of the systems are vital to the users' willingness of adoption and continued use. To that end, Corbett and Le Dantec have presented a design framework for trust in digital civics that seeks to improve community engagement and may be adopted by current and future designers of civic tech [12].

*7.3.3 Closely engage citizens to inform civic tech design.* The importance of engaging citizens in the design and development of civic tech across various stakeholders has been stressed even despite acknowledgements that such efforts require significant time and monetary investments [78,81]. Forms of engagement include expanding the depth and scale of citizen participation in civic activities [66], forming collaborations with researchers such as through participatory action research to glean insights from within the communities [4,68,78], partnering with technical experts to identify meaningful areas for development [81], and communicating with governments to better raise concerns and suggestions to those with the greatest capacity to act upon them [52].

*7.3.4 Expand the scope of investigation in civic tech design.* Areas for improvement in the design of civic tech remain despite the successes these systems have in enhancing public processes such as in facilitating new and alternative forms of social innovation and in better engaging citizens in public concerns [72]. Beyond its utility, negative affordances of civic tech such as inadvertently enabling malice and misuse should be examined and addressed to maintain a safe and healthy environment [8]. Additionally, rather than striving to diminish the needs and deficiencies of the people, designers can instead consider designing for civic tech that builds on the strengths and assets that already exist within the community such as through the asset-based approach proposed by Dickinson et al. [16].

*7.3.5 Get greater government support in civic tech initiatives.* A significant number of civic tech solutions that are created by average citizens are perpetually threatened by a lack of funding and restrictive government legislation. Furthermore, the informality of these systems raises the issue of accountability when things go wrong, as has already been prefaced by social media in the last decade [78]. To this end, there has been calls for greater involvement by the government to support and regulate civic tech initiatives. The support can come in the form of legislation to require civic bodies to intimately engage with communities, or with encouraging entrepreneurship and investment in civic tech [47]. Gastil and Davies have further proposed the notion of a digital democracy through the establishment of the Corporation for Public Software that serves as an independent entity to curate and provide funding for various key actors involved in the creation and maintenance of civic tech that serve public interests [27].

*7.3.6 Work towards what can be.* Finally, civic tech initiatives can look beyond addressing what currently exists to working towards what can be. There has been much work on alternative forms of practices that challenges neoliberalism present in most democracies today. Meng et al. calls for efforts towards a caring democracy, demonstrating that collaborative data work by concerned citizens to enact positive change in the community can occur even on a small scale [54]. In a similar vein, Heitlinger et al. show how citizens can claim urban spaces through the integration of agriculture-related civic tech in a community garden to pave the way for sustainable smart cities [38]. These works advocate for citizens to take ownership of the spaces they inhabit to build more strongly connected and healthier communities through altruistic and sustainable practices, giving a glimpse into the possibilities of a less dystopian future when collective action, even if in pocket-sizes, is taken.

# 8 CONCLUSION

## 8.1 Limitations

This review paper has its limitations. Firstly, we emphasized the presence of a keyword "civic" in the corpus because we are interested in prior studies that at least considered the civic dimension of technological development. It is indeed true that earlier terms such as e-government may be relevant here (e.g., if an e-government service aims to encourage civic engagement). But in order to discern those studies that do not contain a purpose of benefiting the citizens (as part of our definition of civic tech), including "civic" as a keyword is necessary. Nevertheless, we are not attempting to make a final conclusion about civic tech studies. We are aware that other keywords such as "benevolent tech" or tech for "social good" may lead to more relevant papers. Secondly, our search was conducted in the ACM library, limiting us to the ACM-published studies only. Other CHI and CSCW publication venues such as International Journal of Human-Computer Studies may contain similar papers. Moreover, other academic disciplines have also contributed to the field. Political science, communication research, and education research are homes to civic tech studies, too. Future research can expand to include these disciplines and their publication venues such as Journal of Deliberative Democracy (formally known as Journal of Public Deliberation), the EGOV-CeDEM-ePart conference proceedings, and more.

Civic Tech as a research field has reached a stage whereby a rich body of knowledge has been accumulated and a large range of technological tools has been experimented with to serve the benefits of citizens. Previous achievements include an interdisciplinary knowledge base that draws inspirations from social science research, HCI design theories, and studies on cooperative works; an international community that includes both academics and practitioners; and a multi-stakeholder framework of collaboration and coproduction. Our review also reveals that civic issues are long-lasting, complex, and context sensitive. To bring civic tech to the next level, we need stronger commitment from the key stakeholders, such as governments committing to build civic tech into its official system; commercial entities committing to develop mainstream technologies that are in line with civic values; designers committing to actively engage citizens and other civic actors in their design; and citizens committing to educate themselves and devote time and energy to managing civic issues.

**ACKNOWLEDGMENTS**
Blind for review